\documentclass[aps,pra,a4paper,reprint,amsmath,amssymb,graphicx,longbibliography, showkeys, superscriptaddress]{revtex4-1}

\usepackage[english]{babel}
\usepackage[utf8]{inputenc}
\usepackage[colorinlistoftodos, color=green!40, prependcaption]{todonotes}
\usepackage{float} 
\usepackage{xcolor}

\usepackage[unicode]{hyperref}
\hypersetup{
   unicode=true,          % non-Latin characters in Acrobat??s bookmarks
   plainpages=false,
   colorlinks=true,       % false: boxed links; true: colored links
   citecolor=blue,        % color of links to bibliography
}

\usepackage[T1]{fontenc}

\newcommand{\mods}[1]{\left|#1\right|^2}
\DeclareMathOperator{\sech}{sech}

\newcommand{\lJ}{\xi_\mathrm{J}}

\DeclareRobustCommand{\rev}[1]{\textcolor{blue}{#1}}

\begin{document}

\title{Rotational pendulum dynamics of a vortex molecule in a channel geometry}
\author{Sarthak Choudhury}
\affiliation{Dodd-Walls Centre for Photonic and Quantum Technologies, New Zealand Institute for Advanced Study, Centre for Theoretical
Chemistry and Physics, Massey University, Auckland 0632, New Zealand}

\author{Joachim Brand}
\affiliation{Dodd-Walls Centre for Photonic and Quantum Technologies, New Zealand Institute for Advanced Study, Centre for Theoretical
Chemistry and Physics, Massey University, Auckland 0632, New Zealand}

\date{\today} % Leave empty to omit a date xxx

\begin{abstract}
A vortex molecule is a topological excitation in two coherently coupled superfluids consisting of a vortex in each superfluid connected by a domain wall of the relative phase, also known as a Josephson vortex. 
We investigate the dynamics of this excitation in a quasi-two-dimensional geometry with slab or channel boundary conditions using an extended point vortex framework complemented by Gross-Pitaevskii simulations. Apart from translational motion along the channel, the vortex molecule is found to exhibit intriguing internal dynamics including rotation and rotational-pendulum-like dynamics. Trajectories leading to a boundary-induced break-up of the vortex molecule are also described qualitatively by the simplified model. We classify the stable and unstable fixed points as well as separatrices that characterize the vortex molecule dynamics.  
\end{abstract}


\maketitle

%\tableofcontents

% \input{sections/section01.tex}  %I believe leaving the sections in separate files is more organized, change it if you desire 
\section{Introduction} \label{sec:outline}
 Nonlinear topological excitations like vortices have been the topic of study in many fields ranging from high-energy  to condensed-matter
physics \cite{Thouless1998,Manton2004}.
They are long lived and stable due to protection by topological constraints and can only be destroyed by annihilation with opposite charges, or by moving out of the superfluid domain.
Intriguing examples of topological excitations are vortex molecules \cite{Garcia-Ripoll2002,Kasamatsu2004}, which exist in two-component superfluids with linear coupling.

Recent experimental progress has made it possible to study two-dimensional two-component Bose-Einstein condensates (BECs) with homogeneous linear (Rabi) coupling between the two components \cite{Nicklas2015,Farolfi2021,Farolfi2021a}, thus creating an extended linearly-coupled two-component superfluid. The linear coupling tends to align the phases of the two condensates. As such, a vortex filament piercing only one of the two condensates initiates a domain wall of the relative phase \cite{Son2002}, which can terminate at an antivortex in the same condensate, or at a vortex in the other one. The latter situation is referred to as a vortex molecule \cite{Kasamatsu2004}, or sometimes a fractional vortex molecule \cite{etoCollisionDynamicsReactions2020}, since either of the two individual vortices only carries a fraction of the total vortex charge. 
Vortex molecules have been studied extensively in the theoretical literature  \cite{Garcia-Ripoll2002,Kasamatsu2004,Cipriani2013,Tylutki2016,kasamatsuShortrangeIntervortexInteraction2016,Calderaro2017a}. Interest in vortex molecules is partly motivated by the fact that the domain wall creates an energy cost that is approximately linear with the separation of the two vortices, which evokes analogies to color confinement in quantum chromodynamics \cite{Eto2018,etoCollisionDynamicsReactions2020}.

Predicting and understanding vortex dynamics is a challenging problem. A ubiquitous situation in ultracold gas experiments is the elongated or cigar-shaped geometry \cite{Ketterle2008a,Pethick2008,Becker2013}, where a vortex perpendicular to the long trap axis is a stable nonlinear excitation in a scalar superfluid \cite{Mateo2015a}. In such an elongated trap, a single vortex becomes a localised excitation on the length scale of the narrow trap diameter resembling a dark soliton, which gives rise to the concept of a solitonic vortex \cite{brand01a,Brand2002,Komineas2003,Yefsah2013,kuMotionSolitonicVortex2014, Donadello2014,Toikka2016}.
% in having a characteristic phase step across that determines its speed of motion relative to the superfluid. 

In this work we analyse the motion of a vortex molecule in a channel, or slab geometry that is extended in one dimension and has parallel hard-wall boundaries in the second. We assume the third dimension to be tightly confined to the order of the healing length or smaller, such that the problem effectively becomes two-dimensional. This channel geometry embodies the essential qualitative features of the ubiquitous elongated atom trap, while at the same time providing access to analytical treatment. Furthermore, near homogeneous potentials with hard walls, so-called flat bottom traps, have become increasingly available to experiments in recent years  \cite{Chomaz2015,kwonSoundEmissionAnnihilations2021}.

The dynamics of a single vortex in a channel was analysed in Ref.~\cite{Toikka2016} starting from the method of images and applying compressible corrections as a perturbation. For a vortex molecule, the presence of the domain wall connecting the vortices provides an interaction potential, which has an interesting interplay with the effects of the channel boundaries on the vortex motion. Here, we develop a simple model for the dynamics of a vortex molecule augmenting the method of images by a parameterised interaction potential capturing the effects of the domain wall. Similar ideas have previously been implemented to understand the rotation dynamics of a centered vortex molecule in an isotropic harmonic trap \cite{Tylutki2016,Calderaro2017a}.
For the channel geometry the model predicts a rich phase space for the vortex molecule dynamics with different dynamical regimes separated by separatices. A particularly intriguing  rotational-pendulum-like regime of motion is predicted
%In particular 
in the case of repulsive cross-condensate nonlinear interactions where the vortex-vortex interaction has a minimum at finite vortex separation.
%, an intriguing dynamical regime resembling the motion of a rotational pendulum is predicted. 
Numerical simulations with the Gross-Pitaevskii equation (GPE) complement and support the predictions of the simplified model.

The paper is structured as follows. Section \ref{sec:develop} introduces the system in
light of the GPE. Section \ref{sec:model} introduces the main point-vortex model and its equations of motion.
Section \ref{sec:results} discusses the resulting dynamics of the vortex molecule comparing predictions  from the  
point vortex model with full time-dependent simulations of  the GPE dynamics, with conclusions provided in  Sec.~\ref{sec:conclusions}.
Appendix \ref{sec:appendixA} provides  details on the  calculation and the parametrization of the vortex molecule energy and
the  twisted projective plane boundary  conditions used in the calculations.

\section{Mean-Field Formulation} \label{sec:develop}

We describe a system of two linearly coupled Bose-Einstein condensates with complex order parameters
$\psi_{1}(\mathbf{r},t)$ and $\psi_{2}(\mathbf{r},t)$ 
in two spatial dimensions described by the coupled GPEs
% which not only interact with each other through $s$-wave contact interactions but also through a spatially invariant coherent coupling. Within the mean-field framework this is described by a set of coupled GPEs, 
\begin{subequations} \label{coupledGPE}
\begin{align}
% i\hbar\frac{d\psi_i}{dt}=\left(h+g_{12}|\psi_{3-i}|^2 +g_i|\psi_{i}|^2 -\mu \right)\psi_i -\nu \psi_{3-i},
i\hbar\frac{d\psi_1}{dt}&=\left(\hat{h} -\mu +g_1|\psi_{1}|^2 +g_{12}|\psi_{2}|^2\right)\psi_1 -\nu \psi_{2}, \\
i\hbar\frac{d\psi_2}{dt}&=\left(\hat{h} -\mu +g_2|\psi_{2}|^2 +g_{12}|\psi_{1}|^2 \right)\psi_2 -\nu \psi_{1},
\end{align}
\end{subequations}
where  $\hat{h}=-\frac{\hbar^2}{2m}\nabla^2+V_{\mathrm{ext}}$ is the single-particle Hamiltonian for bosons of mass $m$, $V_{\mathrm{ext}}(\mathbf{r})$ is an external potential experienced by both components, and $\mathbf{r} = (x, y)^t$ denotes
the vector of spatial coordinates.
In the following, we assume the external potential to provide hardwall boundaries and otherwise be flat, such that we do not have to carry the external potential explicitly. 
%The order parameter components 
%$\psi_{1/2}(\mathbf{r})$
%denotes the two components of the complex  order parameter,
The chemical potential $\mu$ is used to control the particle number in numerical simulations.
% The constant $\mu$ denotes the chemical potential, and $\nu$ is the strength of a homogeneous coherent coupling.
The coupling constants $g_1$ and $g_2$ describe the intra-component nonlinear interactions, and $g_{12}$ the inter-component nonlinearity. 
%strength. $g_i$ is the $i$th nonlinear s-wave intra-species
%interaction and $g_{12}$ is the inter-species interaction. 
The physics of Eq.~\eqref{coupledGPE}
can be experimentally realized by a BEC of ultracold atoms restricted to two hyperfine states, e.g. $^{23}$Na as in Ref.~\cite{Farolfi2021a} where $g_{12} \approx 0.9 g_1\approx 0.9 g_2$.
A spatially homogeneous coherent (Rabi) coupling between the hyperfine states with the energy scale $\nu$ can be provided by driving a radio-frequency or a two-photon microwave transition continuously. Using different atomic species, such as $^{41}$K may make it possible to tune the cross-component coupling constant $g_{12}$ with a Feshbach resonance \cite{Fialko2015}.
Alternatively, the physics of Eq.~\eqref{coupledGPE} with  $g_{12}=0$ could also be accessed by using a single-component BEC and double-well potential in $z$ direction where barrier tunneling provides the linear coupling $\nu$ and the component order parameters $\psi_{1/2}(\mathbf{r})$ are realised in the different wells \cite{Schweigler2017a}. Ensuring homogeneity in two spatial dimensions will be more difficult with such a setup, however.
% While Eq.~\eqref{coupledGPE} conserves the total particle number, the individual particle numbers per component are not conserved. 
To avoid phase separation, we assume $g_{12}^2 < g_1 g_2$. For simplicity, we choose $g \equiv g_1=g_2>0$ and $\nu>0$ \cite{Brand2010}. The unbalanced case i.e. $g_1\neq g_2$ offers 
additional effects like relative buoyancy between the components and scale separation for the healing length of each component, which have been discussed in the literature \cite{Matthews1999,Perez2000,Jezek2001,Chui2001,Gallemi2018}.
% a rich variety of effects %\cite{Richaud2020,Richaud2021,Gallemi2018} which is not within the scope of this work.

The free energy associated with the GPE~\eqref{coupledGPE} is given by
\begin{align}
      W= &\int \bigg[   \sum_{i=1}^2 \left(\psi_i^*\hat{h}\psi_i  + \frac{g_i}{2}|\psi_i|^4 -\mu |\psi_i|^2 \right) \nonumber \\ 
      &+g_{12}|\psi_1|^2|\psi_2|^2 -\nu (\psi_1^*\psi_2+\psi_1\psi_2^*)  \bigg]  d\mathbf{r}. 
\end{align} 
Numerically we find low energy solutions by propagating Eq.~\eqref{coupledGPE} in imaginary time, i.e.~replacing $t\to -i\tau$, which corresponds to minimizing the free energy $W$ by gradient flow.

A trivial  or ground state solution of Eq.~\eqref{coupledGPE}  (for $V_{\mathrm{ext}}=0$) is found with constant $\psi_1 = \psi_2$, where 
%In a homogenous condensate where $V_{\mathrm{ext}}=0$, 
the densities of the individual component condensates are homogeneous and identical, with 
$\mods{\psi_i} = n_0 \equiv ({\mu+\nu})/({g+g_{12}})$ for $i = 1,2$. %\in\{1,2\}$, 
The healing length $\xi={\hbar}/{\sqrt{m(\mu+\nu)}}$ provides the length scale on which this homogeneous solution is recovered away from forced local inhomogeneities due to solitons, vortices, or boundary conditions.

Vortex molecules are composed of a vortex in each component connected by a domain wall of the relative phase.
Relevant analytically known solution of Eq.~\eqref{coupledGPE} with nonlinear defects are the simple vortex and the Josephson vortex. 

\subsection{Simple vortex} \label{sec:simplevortex}
The simple vortex solution is one where a vortex penetrates both components at the same place. It can be understood of a special case of a vortex molecule where the two vortices occur at the same location. 
To find the solution we assume $\psi_1(\mathbf{r})=\psi_2(\mathbf{r})$, which simplifies Eq.~\eqref{coupledGPE} to the single-component GPE
\begin{align}
i\hbar\frac{d\psi_1}{dt}&=\left(\hat{h} -\mu_\mathrm{eff} +g_\mathrm{eff}|\psi_{1}|^2\right)\psi_1,
\end{align}
with $\mu_\mathrm{eff} = \mu+\nu$ and $g_\mathrm{eff}=g_1+g_{12}$, the vortex solutions of which are well known on an infinite domain \cite{Pitaevskii2016a}. They are characterised by a singular phase distribution, an integer vortex charge $\kappa$, and a density node at the vortex location. Specifically for a vortex located at the origin of the coordinate system, 
\begin{align} \label{eq:simplevortex}
\psi_1(\mathbf{r}) = \psi_2(\mathbf{r}) = \sqrt{n_0} f_\kappa(r/\xi) e^{i\kappa \phi},
\end{align}
where $(r, \phi)$ are the polar coordinates, and $f_\kappa$ is a dimensionless function with $f_\kappa(0)=0$ (for $\kappa \neq 0$) and $f_\kappa(\infty)=1$ \cite{Pitaevskii2016a}. Due to phase gradients that decay only weakly away from the vortex singularity, the excitation energy of the simple vortex solution diverges logarithmically with the integration domain.

\subsection{Josephson vortex}
The Josephson vortex is a stationary solution of the coupled GPEs~\eqref{coupledGPE} that realises a domain wall of the relative phase. 
The solution exists for $\nu <\mu/3$ and is homogeneous in one dimension (say along the $y$ coordinate) and inhomogeneous in the other \cite{Kaurov2005, Kaurov2006}
\begin{align}
\label{1dJv}
\psi_{1/2}(\mathbf{r}) =& \sqrt{n_0}  \left[\tanh\left(\frac{x}{\lJ}\right)  \pm i \sqrt{\frac{\mu-3\nu}{\mu+\nu}} \sech\left(\frac{x}{\lJ}\right)  \right] ,
\end{align}  
where $\lJ=\hbar/\sqrt{4m\nu}$ is the Josephson vortex length scale. The stationary Josephson vortex is connected to a single-parameter family of moving solitary-wave solutions, which were characterized in Ref.~\cite{Shamailov2018}. The whole family of solutions is dynamically stable in one spatial dimension (corresponding to tight confinement in the $y$ dimension) For   $\nu \lessapprox 0.15 \mu$ the stationary Josephson vortex is a local minimum of the dispersion relation. This means that it has a positive effective mass \cite{Shamailov2018} and thus is dynamically stable also in two dimensions \cite{Kamchatnov2008,Gallemi2019,Ihara2019}. For $0.15 \mu\lessapprox \nu < \mu/3$ the Josephson vortex has negative effective mass and suffers the snaking instability with eventual decay into vortices similar to the instability of dark solitons \cite{Muryshev1999,Brand2002}. At $\nu \to \mu/3$ the Josephson vortex solution reaches a bifurcation point where it becomes identical to a dark soliton \cite{Kaurov2005,Shamailov2018}.

The energy (line) density of the Josephson vortex is 
\begin{align} \label{eq:sigmaJ}
\sigma_\mathrm{J} = \frac{W_\mathrm{JV} - W_\mathrm{hom}}{L} = \frac{8\hbar \sqrt{\nu}}{3\sqrt{m}} \frac{3\mu-\nu}{g+g_{12}},
%= \frac{8\hbar \sqrt{\nu}(3\mu-\nu)}{3\sqrt{m}(g+g_{12})},
\end{align}
where $W_\mathrm{JV}$ and $W_\mathrm{hom}$ are the free energies of the Josephson vortex and the homogeneous solution, respectively, and $L$ is the extent of the integration domain in the $y$ direction. Approximate descriptions of a domain wall of the relative phase as a soliton solution of the sine-Gordon equation are sometimes used \cite{Son2002,Kaurov2005}. These reproduce the properties of the Josephson vortex solutions of the GPE, including the energy, to leading order in $\sqrt{\nu/\mu}$, i.e.~when the linear coupling is a small parameter \cite{Shamailov2018}.

\subsection{Vortex molecule}

In this work we are interested in the dynamics of a vortex molecule in a channel geometry. Thus we consider a channel of width $D$ aligned along the $x$-axis with hard wall boundaries at $|y|\ge D/2$. We further use a finite computational domain with $x\in (-D, D]$ with antiperiodic boundary conditions, i.e.\ adding a $\pi$ phase to each of $\psi_{1/2}$, as appropriate for a single vortex.

In order to obtain a vortex molecule numerically, we imprint the known phase profile of a vortex in a channel for a single incompressible superfluid \cite{Toikka2016} with different vortex positions in each component and then evolve to a low-energy configuration using imaginary-time evolution. Imaginary time evolution quickly removes most excitations but is slow to move vortex singularities. While local minima of the free energy $W$ are obtained by evolving in imaginary time until convergence, evolution for a finite amount of imaginary time will yield near ideal field configurations corresponding to the lowest energy for the given position of the vortex singularities. Vortex positions are located by accurately tracking the positions of the phase singularities using the software library \texttt{VortexDistributions.jl} \cite{Bradley2022}. 

\begin{figure}
\centering
\includegraphics[width=1\linewidth]{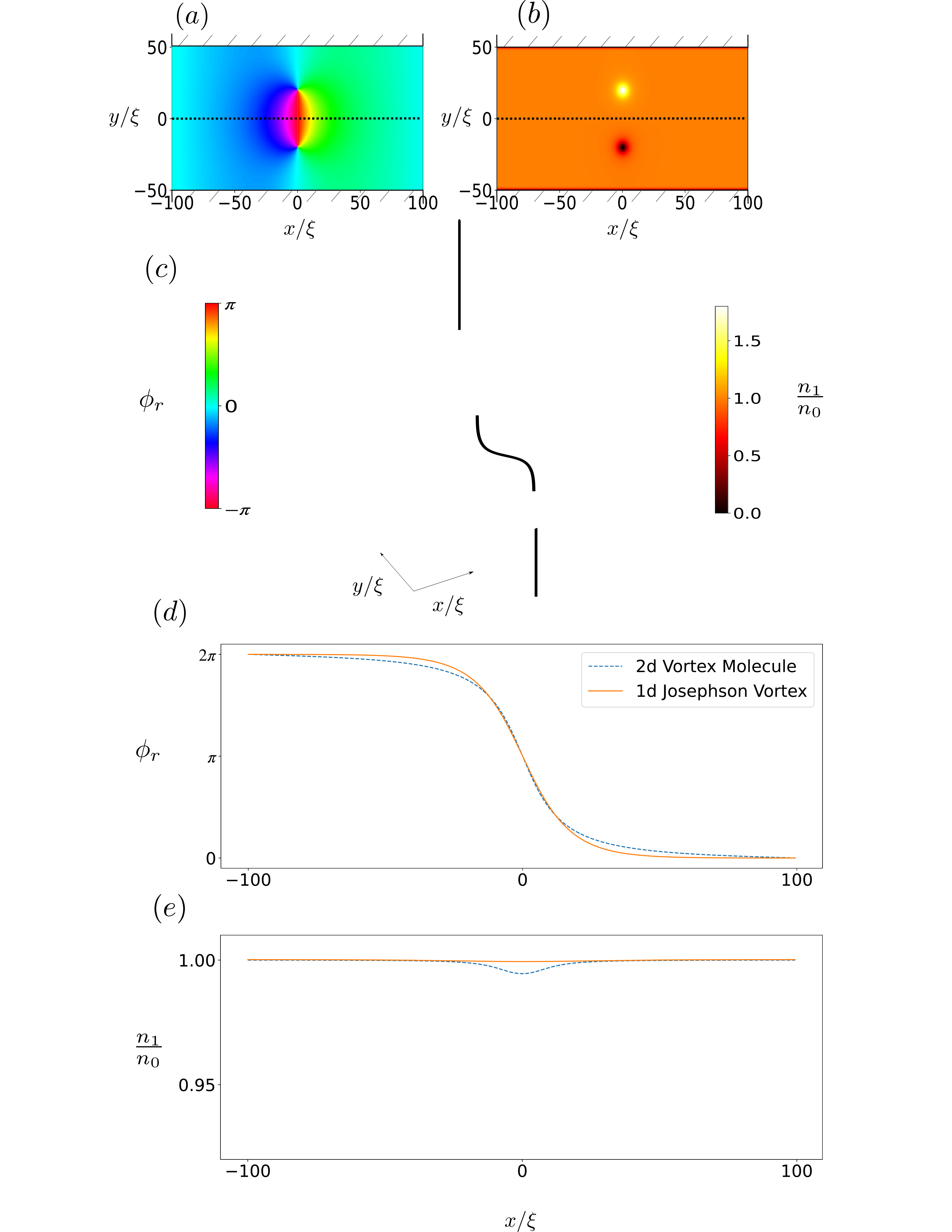}
\caption{
Vortex molecule in a channel geometry of width $D=100\xi$ with hard-wall boundaries at $y=\pm D/2$.
(a) Relative phase $\phi_r(\mathbf{r})=\arg(\psi_1\psi_2^*)$ 
showing the domain wall oriented along the $y$ axis with a large phase gradient. The singular termination points indicate the locations of vortices in the component condensates. 
(b) Single component density $n_1(\mathbf{r})= |\psi_1|^2$ with a density depletion and local density maximum at the locations of the vortices in components 1 and 2, respectively.
%The dotted line at $y=0$ indicated the location of the cut shown in panel 
(c) Concept diagram showing how a vortex filament can be understood to thread the two-component condensate. The density of component 1 is shown on the lower plane and that of component 2 is shown on the upper plane.
(d) Relative phase from panel (a) along the $y=0$ line [dotted line in panel (a)], shown as the dashed blue line in comparison to the relative phase $\arg(\psi_1\psi_2^*)$ from the analytic Josephson vortex solution of Eq.~\eqref{1dJv} (full red line).
(e) Density of component 1 from panel (b)  along the $y=0$ line [dotted line in panel (b)], shown as the dashed blue line in comparison to the component density $|\psi_1|^2$ from the analytic Josephson vortex solution of Eq.~\eqref{1dJv} (full red line).
%(b) Single component density $n_1(\mathbf{r})= |\psi_1|^2$ with a density depletion and local density maximum at the locations of the vortices in components 1 and 2, respectively.
%(c) Concept diagram showing how a vortex filament can be understood to thread the two-component condensate. The density of component 1 is shown on the lower plane and that of component 2 is shown on the upper plane. 
%(d) 
%%
%%in a channel geometry. The phase between the vortex molecule changes by 2$\pi$ and is characteristic of a Josephson vortex.  \\
%(b) Density of one condensate  $n_1=|\psi_1/\psi_0|^2$ where $\psi_0=\sqrt{(\mu+\nu)/(g+g_{12})}$. The vortex is the region with very low density while the 
%region of high density is the position of the other vortex in the other condensate. \\
%(c) Concept diagram of re-imagining of the vortex molecule as a vortex filament piercing through one condensate and then traveling transversely in the space between the condensates and piercing through the other. \\
%(d) Cross section of (a)  along the middle of the channel along the black line.
%(e) Cross section of (b) similar to (d). The GPE calculations have been compared with the relative phase and density of the 1d analytical form of the Josephson vortex given in Eq.~\eqref{1dJv} in (d) and (e). 
Parameters are $\nu=2\times 10^{-4} \mu$,  $g_{12}=0.9 g$, and $g=0.53 \mu\xi^2$. } 
\label{vfilament}
\end{figure} 

Figure \ref{vfilament} shows a vortex molecule in a channel with width $D=100\xi$. The vortices located in component 1 and 2 show up in the relative phase $\phi_r = \arg(\psi_1 \psi_2^*)$ as vortex and antivortex, respectively, see Fig.~\ref{vfilament} (a). In the single-component density shown in panel (b) the depleted vortex core appears black while the vortex in the other component leads to a local density maximum due to the repulsive cross-component nonlinearity and thus appears as a bright spot. 
Panel (c) shows a three-dimensional schematic indicating how a vortex filament can be understood to thread the arrangement. If the linear coupling between the two components originates from a double-well trap, the components will be displaced in the $z$ dimension as shown. If, on the other hand, Rabi coupling of internal states is used, the separation is merely conceptual. Since the vortex line cannot simply terminate, it must thread between the component, which gives rise to the domain wall of the relative phase. The domain wall structure is clearly seen in the relative phase in Fig.~\ref{vfilament} (a). More detailed views are shown in 
panels (d) and (e), which compare the cross sections of the relative phase and single-component density from panels (a) and (b), respectively, with the exact Josephson vortex solution of Eq.~\eqref{1dJv}. While small differences exist, it is seen that the Josephson vortex solution provides a reasonable description of the domain wall in the vortex molecule.

The domain wall has an energy content, which may be expected to be linear in its length $d$ and approximated by $\sigma_\mathrm{J} d$, 
%jb: Note that sigma_J is an energy line density. The energy content of the domain wall is the product with the domain wall length $d$. 
according to Eq.~\eqref{eq:sigmaJ}. If the domain wall is stretched beyond a critical length, it becomes energetically favourable to generate vortex-antivortex pairs and break up the domain wall into shorter segments \cite{Ihara2019}. Within the picture of Fig.~\ref{vfilament} (c) this can be understood as the vortex filament looping outside of the condensates (or the in-between region), where its existence comes without an energy cost.

\begin{figure}[H]
\centering
\includegraphics[width=1\linewidth]{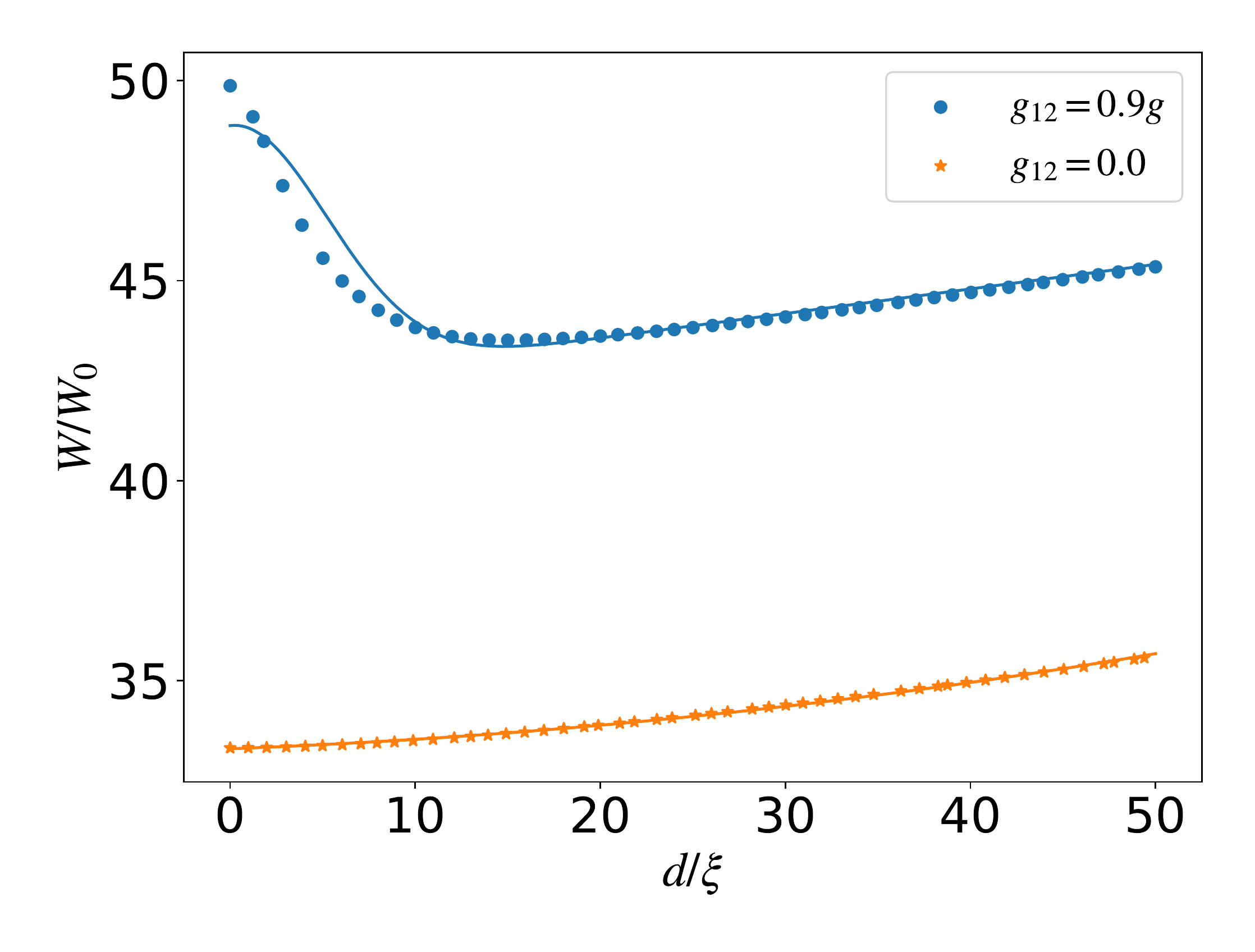}
\caption{Total energy of the vortex molecule as a function of the molecular distance $d$ on a square domain of $180\xi \times 180\xi$. The symbols are numerical results from imaginary time evolution 
and the lines are fits to the numerical data. In the absence of cross-component interactions, the vortex molecule energy is monotonous with a minimum at $d=0$, while for repulsive cross-component interactions at $g_{12} = 0.9g$ and energy minimum appears at the equilibrium distance $d_\mathrm{eq}=14.8 \xi$. Other parameters are $\nu=2\times 10^{-4} \mu$, $g=0.53\mu \xi^2$ for $g_{12}=0.9g$, and 
$g=\mu \xi^2$ for $g=0$. The unit of energy is $W_0={\hbar^2(\mu+\nu)}/{m(g_1+g_c)}$. 
Details of the fitting procedure and boundary conditions are described in Appendix \ref{sec:appendixA}.
} \label{Interaction_parameter}
\end{figure}

In order to better understand and quantify the energy cost of the domain wall, i.e.~the interaction energy of a vortex molecule, we compute the total energy as a function of the molecular size $d$, the distance between the two constituent vortices as shown in Fig.~\ref{Interaction_parameter}. 
%
% Using twisted real-projective-plane boundary condition as described in Appendix  \ref{sec:appendixA} we imprint each condensate with a single vortex phase mask at a distance $\pm d_{\mathrm{ext}}/2$ from the center. Then, we put pinning potentials (peaked Gaussian potentials) on the positions of the phase singularity of each vortex  and evolve the system according to Eq.~\eqref{coupledGPE} in imaginary time. This creates a good starting position for a vortex molecule with initial molecular length $d_{\mathrm{ext}}$. 
%
The computational details and the boundary conditions, which are designed to approximate the vortex molecule on an infinite plane, are described in Appendix  \ref{sec:appendixA}. Results for two different values of the inter-component nonlinear coupling $g_{12}$ are shown, and neither is strictly linear, indicating that other effects come into play in addition to the linear domain wall contribution. Moreover, the slope is consistently less than the Josephson vortex energy density $\sigma_\mathrm{J}$ consistent with a finding of Ref.~\cite{Eto2018}.
For $g_{12}=0$ the energy is monotonous as a function of $d$, and the lowest energy configuration is at $d=0$, i.e.\ when the vortex molecule realises the simple vortex solution of Eq.~\eqref{eq:simplevortex}. When $g_{12} > 0$ the repulsive inter-component nonlinearity favors filling the vortex core in one component with density from the other, which leads to an energy benefit when the vortex cores do not overlap. In this case the energy has an energy minimum at a finite molecular distance, which becomes a stable equilibrium of the vortex molecule in real-time evolution.

\section{Extended Point-vortex Model} \label{sec:model}

The potentially complicated dynamics of a condensate described by the GPE, a partial differential equation, can be simplified considerably by reducing it to the motion of point vortices. This is justified when no or little other excitations such as solitons or phonon radiation are present or generated, i.e.~when the motion proceeds by moving near adiabatically through low-energy vortex configurations. In this case the motion can be described in a Hamiltonian framework just from knowing the energy (gradients) of the different vortex configurations \cite{Newton2001}. In the case of a near-homogeneous BEC with hard-wall boundary conditions this is greatly aided by the method of images.
The method of images is exact for an incompressible and irrotational fluid, and becomes a useful approximation for the GPE on length scales large compared to the healing length. 
Here we combine the numerically determined interaction energy of a vortex molecule with the method of images for capturing the influence of the channel boundaries on the vortex motion.

\subsection{Single component vortex in a channel} \label{sec:singlevortex}

Reference \cite{Toikka2016} solved the  vortex in a channel in a single-component BEC starting from the method of images and developing compressible corrections as a power series in $(\xi/D)^2$. We summarise some of the results and use them as a starting point. Ignoring the compressible corrections and a constant offset, the energy of a single vortex in a channel extended along the $x$ direction with walls located at $y = \pm D/2$ is
\begin{align} \label{eq:ESV}
E_\mathrm{SV}(Y) = \frac{\pi \hbar^2 \kappa^2 n}{m} \ln \cos \left(\frac{Y}{D}\pi\right) ,
\end{align}
where $n$ is the (background) density and $Y$ is the $y$-displacement of the vortex from the origin (with $-D/2 < Y < D/2$).
The velocity field (phase gradient) of the vortex solution is exponentially localised in the $x$ dimension on the length scale $D$. 
The momentum in $x$ direction is simply proportional to $Y$,
\begin{align}
{P_\mathrm{SV}} = 2\pi n\hbar \kappa Y ,
\end{align}
which is consistent with the phase space for vortex motion being two-dimensional. 

Following Ref.~\cite{Newton2001}, it is convenient to introduce a rescaled Hamiltonian function
\begin{align}
\mathcal{H}(X,Y) = \frac{E(X, Y)}{2\pi n \hbar \kappa} ,
\end{align}
where $E(X, Y)$ is the energy of a vortex with coordinates $X$ and $Y$.
With this definition, the $y$ coordinate of a vortex becomes the canonical momentum of its $x$ coordinate, and Hamilton's equations take the form
\begin{subequations}
\begin{align}
\dot{X} &= \frac{\partial \mathcal{H}}{\partial Y} ,\\
\dot{Y} &= -\frac{\partial \mathcal{H}}{\partial X}  .
\end{align}
\end{subequations}

For the single vortex in the channel, we find [with $E(X, Y) = E_\mathrm{SV}(Y)$]
\begin{subequations} \label{eq:singlevortexeom}
\begin{align}
\dot{X}& = -\frac{\pi \kappa \hbar}{2mD} \tan  \left(\frac{Y}{D}\pi\right) ,\\
\dot{Y} &= 0 .
\end{align}
\end{subequations}
A single vortex 
thus propagates at constant velocity along the channel, 
i.e.~in the $x$ direction. 
The velocity depends on the (constant) $Y$ position in the channel. It vanishes when the vortex is situated in the center of the channel (at $Y=0$) and diverges as the vortex molecule approaches the edges of the channel.  Note that this divergence is regularized and disappears for a compressible BEC as the predictions from the point vortex model become invalid when the vortex separation from the boundaries is less than the healing length $\xi$.
The effective mass is given by \cite{Toikka2016}
\begin{align}
{M_\mathrm{SV}} &=  \frac{d{P_\mathrm{SV}}}{d\dot{X}} = {\left( \frac{\partial^2\mathcal{H}}{\partial P_\mathrm{SV}^2}\right)^{-1} = \frac{(2\pi n\hbar \kappa)^2}{E_\mathrm{SV}''(Y)}
}
\nonumber \\
&=  -\frac{4}{\pi} m n D^2 \left[\cos \left(\frac{Y}{D}\pi\right)\right]^2 .
\end{align}
It is negative and its magnitude is approximately 
the mass of the superfluid enclosed by the area $D^2$ while the vortex is near the center of the channel.
 
\subsection{Vortex molecule point vortex model} \label{sec:vmPointVortexModel}

For the Hamiltonian of the vortex molecule we use a simple ansatz where we simply add the energies of each vortex in the channel and an interaction energy
\begin{align}\label{eq:HVM}
\mathcal{H}_\mathrm{VM}(X_1,X_2, Y_1,Y_2) =
\frac{E_\mathrm{SV}(Y_1) + E_\mathrm{SV}(Y_2) + V(d)}{2\pi n \hbar \kappa} ,
\end{align}
where $V(d)$ is an interaction energy that depends only on the distance $d = \sqrt{(X_1-X_2)^2 + (Y_1-Y_2)^2}$
between the two vortices.
The equations of motion then become
\begin{subequations}\label{eq:eomVM}
\begin{align}
\dot{X}_{1/2} &= \frac{\partial \mathcal{H}_\mathrm{VM}}{\partial Y_{1/2}} ,\\
\dot{Y}_{1/2} &=  -\frac{\partial \mathcal{H}_\mathrm{VM}}{\partial X_{1/2}} .
\end{align}
\end{subequations}

The phase space of the vortex molecule is four dimensional and more complex than that of a single vortex in a channel. 
While the motion of the center of mass does not fully decouple from the relative motion, it still does so approximately when the center of mass is close to the center of the channel. In particular, when the molecule is symmetrically centered in the channel with $Y_1 = - Y_2$ then it follows from Eqs.~\eqref{eq:eomVM}  and \eqref{eq:HVM} and the fact that  $E_\mathrm{SV}(Y)$ of Eq.~\eqref{eq:ESV} is an even function of $Y$, that 
$\dot X_1+ \dot X_2 = 0 = \dot Y_1+\dot Y_2$. I.e.~the center of mass is stationary and the phase space of the vortex molecule motion reduces to the two-dimensional phase space of relative motion. 

\subsection{Approximate separation of the center-of-mass motion} \label{sec:comseparation}

In order to obtain more insights
we introduce a symmetric transformation to new canonical coordinates for center-of-mass ($\tilde{Q}, \tilde{P}$) and relative motion ($\tilde{q}, \tilde{p}$)
\begin{subequations}
\begin{align}
\tilde{q} &= \frac{X_1 - X_2}{\sqrt{2}}, & \tilde{Q} &=\frac{X_1 + X_2}{\sqrt{2}}, \\
\tilde{p} &= \frac{Y_1 - Y_2}{\sqrt{2}}, & \tilde{P} &=\frac{Y_1 + Y_2}{\sqrt{2}}.
\end{align}
\end{subequations}
The Hamiltonian function in the new coordinates is
\begin{align}
\tilde{\mathcal{H}}(\tilde{q},\tilde{Q},\tilde{p},\tilde{P}) &=\mathcal{H}_\mathrm{VM}(\frac{\tilde{Q} + \tilde{q}}{\sqrt{2}},
\frac{\tilde{Q} - \tilde{q}}{\sqrt{2}},\frac{\tilde{P} + \tilde{p}}{\sqrt{2}},\frac{\tilde{P} - \tilde{p}}{\sqrt{2}}),
\end{align}
with $\mathcal{H}_\mathrm{VM}$ given by Eq.~\eqref{eq:HVM}. By expansion of the relevant terms in powers of $\tilde{P}$ and $\tilde{p}$ we find
that the Hamiltonian can be written in the approximately separable form
\begin{align}
\tilde{\mathcal{H}}(\tilde{q},\tilde{Q},\tilde{p},\tilde{P}) &= \tilde{\mathcal{H}}_\mathrm{com}(\tilde{Q},\tilde{P}) 
+  \tilde{\mathcal{H}}_\mathrm{rel}(\tilde{q},\tilde{p})  + \mathcal{O}(\tilde{P}^2\tilde{p}^2) ,
\end{align}
which confirms that relative motion can be considered independently at or close to a fixed point of the center-of-mass motion with $\tilde{P}=0$, consistent with the result from the previous section. Conversely, center-of-mass motion can be considered independently at a fixed point of the relative motion with $\tilde{p}=0$.
The center-of-mass motion described by
\begin{align} \label{eq:Hcom}
\tilde{\mathcal{H}}_\mathrm{com}(\tilde{Q},\tilde{P})  =  \frac{E_\mathrm{SV}({\tilde{P}}/{\sqrt{2}})}{\pi n \hbar \kappa},
\end{align}
which is, up to rescaling factors, that of a single-component vortex in a channel.
Displacement from center in the $y$ direction thus induces a constant velocity in $x$ direction according to Eq.~\eqref{eq:singlevortexeom}.
The center-of-mass effective mass in physical units is
\begin{align}
M_\mathrm{VM} &= 4\pi n \hbar \kappa \left(\frac{\partial^2 \tilde{\mathcal{H}}_\mathrm{com}}{\partial \tilde{P}^2}\right)^{-1} \nonumber \\
&= 2 M_\mathrm{SV}({\tilde{P}}/{\sqrt{2}}) ,
\end{align}
\rev{which} is twice the mass of a single vortex in this approximation,
and ${\tilde{P}}/{\sqrt{2}} = ({Y_1+Y_2})/{2}$ is the $y$ position of the vortex molecule's center.
The relative motion is described by
\begin{align}\label{eq:Hrel}
 \tilde{\mathcal{H}}_\mathrm{rel}(\tilde{q},\tilde{p}) = \frac{2 E_\mathrm{SV}({\tilde{p}}/{\sqrt{2}}) 
 + V(\sqrt{2}\sqrt{\tilde{q}^2+\tilde{p}^2})}{2\pi n \hbar \kappa} ,
\end{align}
which captures both the effects of the channel boundary conditions via $E_\mathrm{SV}$ and the molecular interaction via the vortex molecule energy $V$.

\section{Vortex molecule dynamics with fixed center of mass} 
\label{sec:results}

The extended point vortex model of the previous section presents a simple model of vortex motion in a Hamiltonian framework. It greatly reduces the complexity associated with the partial differential equations of the GPE description. Our goal is to show that it can capture the major qualitative features of vortex molecule dynamics appropriate to a given trap geometry with a parameterized vortex interaction.

We consider the dynamics of a vortex molecule in a channel of width $D$ in $y$ direction and infinite extent in $x$ direction. To emulate the infinite channel in our numerical GPE simulations, we use a computational domain of $2D \times D$ extent with hard wall boundaries in $y$ and antiperiodic boundary conditions (periodic with a $\pi$ phase twist) in $x$ direction, which realizes a ribbon with a periodic vortex -- anti-vortex train. Due to the exponential localization of the solitonic vortex (Sec.~\ref{sec:singlevortex} and Ref.~\cite{Toikka2016}), the phase gradients become negligible near the $x$ boundaries, and the single vortex in an infinite channel is well emulated. 

In the extended point vortex model, where energy is conserved,
the trajectories of a vortex molecule are the contour lines of the Hamiltonian $\mathcal{H}_\mathrm{VM}$ in the four-dimensional phase space. 
For the interaction energy $V(d)$, we use a parameterized fit of the total energy of a vortex obtained from imaginary-time evolution in real projective plane boundary conditions, which mimic a vortex molecule on an infinite plane. For details see Appendix \ref{sec:appendixA} and Fig.~\ref{Interaction_parameter}.
In the following we consider the situation where the vortex molecule is aligned symmetrically in the channel and hence its center of mass remains stationary (see Sec.~\ref{sec:vmPointVortexModel}). In this case the dynamics of the vortex molecule is fully captured by the relative motion Hamiltonian $\tilde{\mathcal{H}}_\mathrm{rel}$ of Eq.~\eqref{eq:Hrel}.

\begin{figure*}
\centering
\includegraphics[width=\linewidth]{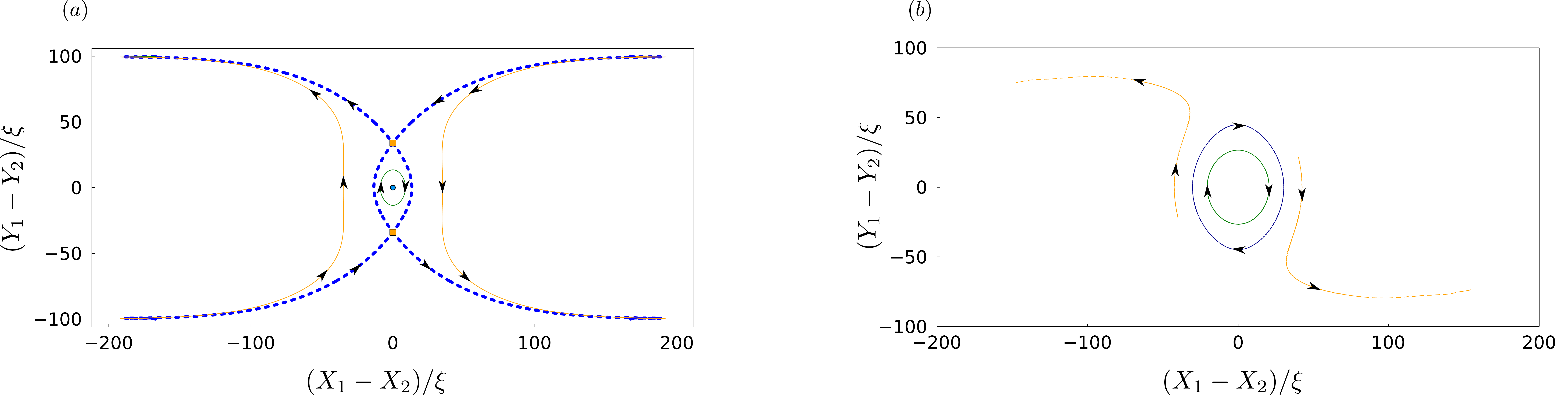}
\caption{Phase space of relative motion for the vortex molecule in a channel of width $D=100\xi$ in the absence of inter-component nonlinearity ($g_{12}=0$). Trajectories are shown as lines with arrows in the relative coordinates of the vortices 
{$X_1 - X_2=\sqrt{2}\tilde{q}$} and 
{$Y_1 - Y_2=\sqrt{2}\tilde{p}$} with a hard-wall boundary at $Y_1 - Y_2=\pm D$. 
(a) Extended point vortex model of Eqs.~\eqref{eq:HVM} and \eqref{eq:eomVM}. Symbols indicate fixed points. The  round blue dot indicates a stable (elliptical) fixed point that is also a local energy minimum. The orange squares indicate saddle points (hyperbolic fixed points). The associated stable/unstable manifolds (dotted blue lines) provide separatrices separating bounded and unbounded motion. 
(b) Vortex trajectories obtained from solving the time-dependent GPE \eqref{coupledGPE}.
Other parameters are $\nu=2\times 10^{-4} \mu$ and $g= \mu\xi^2$.
}  \label{fig:phase_space_g12_0}
 \end{figure*}

The phase space of a vortex molecule in a channel in the absence of inter-component interactions is show in Fig.~\ref{fig:phase_space_g12_0}. The phase space portrait from the point-vortex model in panel (a) is contrasted by the vortex trajectories obtained from GPE simulation in panel (b) with low-energy starting configurations cleaned by imaginary-time evolution.
The central local energy minimum [marked with a blue dot in panel (a)] corresponds to a simple vortex of Sec.~\ref{sec:simplevortex} located in the center of the channel. It is an elliptic fixed point, and the surrounding elliptic trajectories describe the vortex molecule rotating clockwise around its center of mass. A separatrix (dotted blue line) separates the bounded periodic motion from unbounded trajectories where vortices move mainly under the influence of the boundary-induced image vortices. The yellow marked trajectories correspond to motion where vortices in component 1 and 2 approach each other along the channel boundaries, then perform a partial molecule rotation before they move away from each other along the boundary.

\begin{figure*}
\centering
\includegraphics[width=\linewidth]{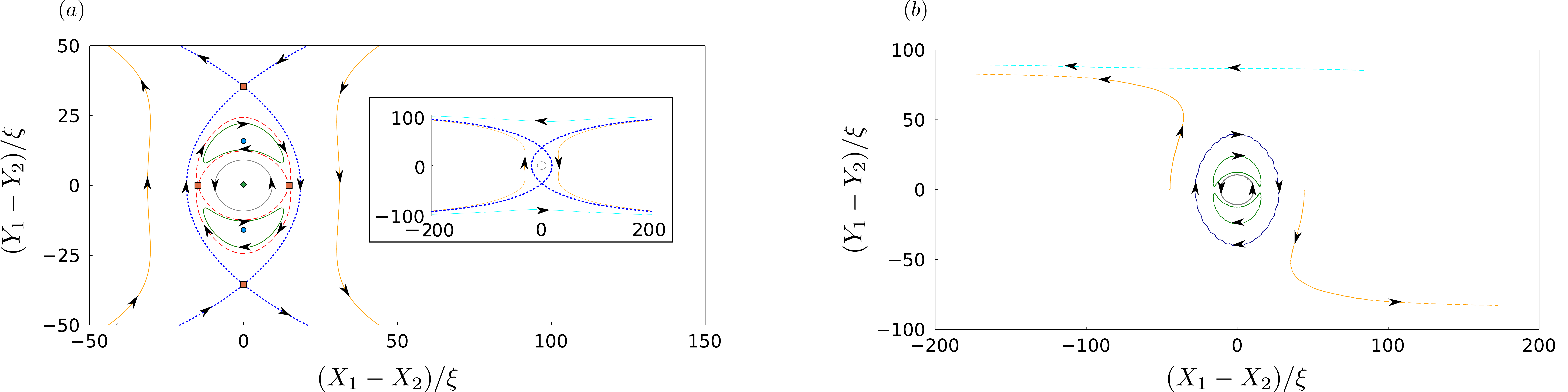}
\caption{Phase space of relative motion for the vortex molecule in a channel  of width $D=100\xi$ in the presence of inter-component nonlinearity ($g_{12}=0.9g$). Trajectories are shown as lines with arrows in the relative coordinates of the vortices 
{$X_1 - X_2=\sqrt{2}\tilde{q}$} and 
{$Y_1 - Y_2=\sqrt{2}\tilde{p}$} with a hard-wall boundary at ${Y_1 - Y_2=}\pm D$. 
(a) Extended point vortex model of Eqs.~\eqref{eq:HVM} and \eqref{eq:eomVM}. Symbols indicate fixed points. The  round blue dots (local energy minima) and the green diamond (local energy maximum) are elliptical fixed points. The orange squares indicate saddle points (hyperbolic fixed points), which give rise to two different sets of disconnected separatrices (blue dotted lines and dashed orange lines).
The inset provides an overview up to the channel boundaries.
(b) Vortex trajectories obtained from solving the time-dependent GPE \eqref{coupledGPE}.
Other parameters are $\nu=2\times 10^{-4} \mu$ and $g=0.53 \mu\xi^2$.
}     \label{fig:phase_space_g12}
\end{figure*}

When a repulsive inter-component interaction of $g_{12}=0.9 g$ is present, the picture changes qualitatively, and the phase space becomes considerably more complex. This is seen in Fig.~\ref{fig:phase_space_g12}. While the dotted (blue) separatrix system with its hyperbolic fixed points stays in place, and outside the phase space remains qualitatively unchanged, the inner domain enclosed by the dotted (blue) separatrix looks very different. 
Instead of a  basin with a single minimum, a distorted Mexican hat shape emerges.
Specifically, the central elliptic fixed point that corresponds to the simple vortex configuration [marked with a green diamond in panel (a)] now marks a local energy maximum. This is due to the energy benefit of off-setting the vortices when the cross-component interaction is repulsive, as already seen in Fig.~\ref{Interaction_parameter}.
As a consequence, the elliptic trajectories surrounding the fixed point have an anti-clockwise orientation in Fig.~\ref{fig:phase_space_g12} (a) and (b). The rim of the Mexican hat is distorted by the effect of the channel boundaries  through $E_\mathrm{SV}(Y)$.
Local energy minima now appear above and below the central fixed point and are marked with blue round dots in panel (a). 
Saddle points with {$Y_1-Y_2=0$} provide hyperbolic fixed points [marked with red squares in panel (a)] and give rise to a new set of 
 separatrices marked with dashed (red) lines.

Due to the changed phase-space structure, we now find crescent shaped trajectories (marked with green lines) that exhibit a rocking motion enclosing the local minima, reminiscent of a rotational pendulum. These trajectories appear close to the equilibrium separation of a vortex molecule in the absence of boundaries seen in Fig.~\ref{Interaction_parameter}. For smaller and larger separations, trajectories showing anti-clockwise and clockwise rotational motion, respectively, are now possible. 

At higher energies,  non-compact vortex trajectories are predicted and observed in both scenarios of Figs.~\ref{fig:phase_space_g12_0} and \ref{fig:phase_space_g12}, where they are marked in yellow and cyan colors. For these trajectories the vortex separation $d$ becomes arbitrarily large, i.e.~the vortex molecule is stretched indefinitely.  Within the point-vortex model, the vortex interaction energy $V(d)$ is assumed to derive from the contribution of a domain wall that extends in a straight line between the two vortices. For the non-compact trajectories, this energy grows without bounds as $d$ increases. This is compensated for by negative energy contributions from $E_\mathrm{SV}$ of Eq.~\eqref{eq:ESV}, which diverges logarithmically as a vortex nears the channel boundary. 

The non-compact trajectories are interesting, because at some point the vortex interaction energy $V(d)$ will be large enough to account for the creation of a vortex-antivortex pair. Such a pair production of vortices could lead to lowering the total energy, as the vortex filament could be threaded outside the coupled superfluid without energy cost, and thus break the linear dependence of the vortex energy  on the separation $d$. Quantum, thermal, or other technical fluctuations are necessary to initiate the pair production because there is an energy barrier to overcome.

The GPE simulations are generally found to follow the predictions of the point vortex model. Animations of the GPE real-time evolution are available in the Supplementary Information for trajectories corresponding to rotational-pendulum-like motion, vortex-molecule rotation, and unbounded motion \cite{SI}. In addition to the vortices following the characteristic trajectories, small amounts of noise originating from radiation due to vortex acceleration are seen there as well \cite{Parker2004a}.

\begin{figure}
\centering
\includegraphics[width=1\linewidth]{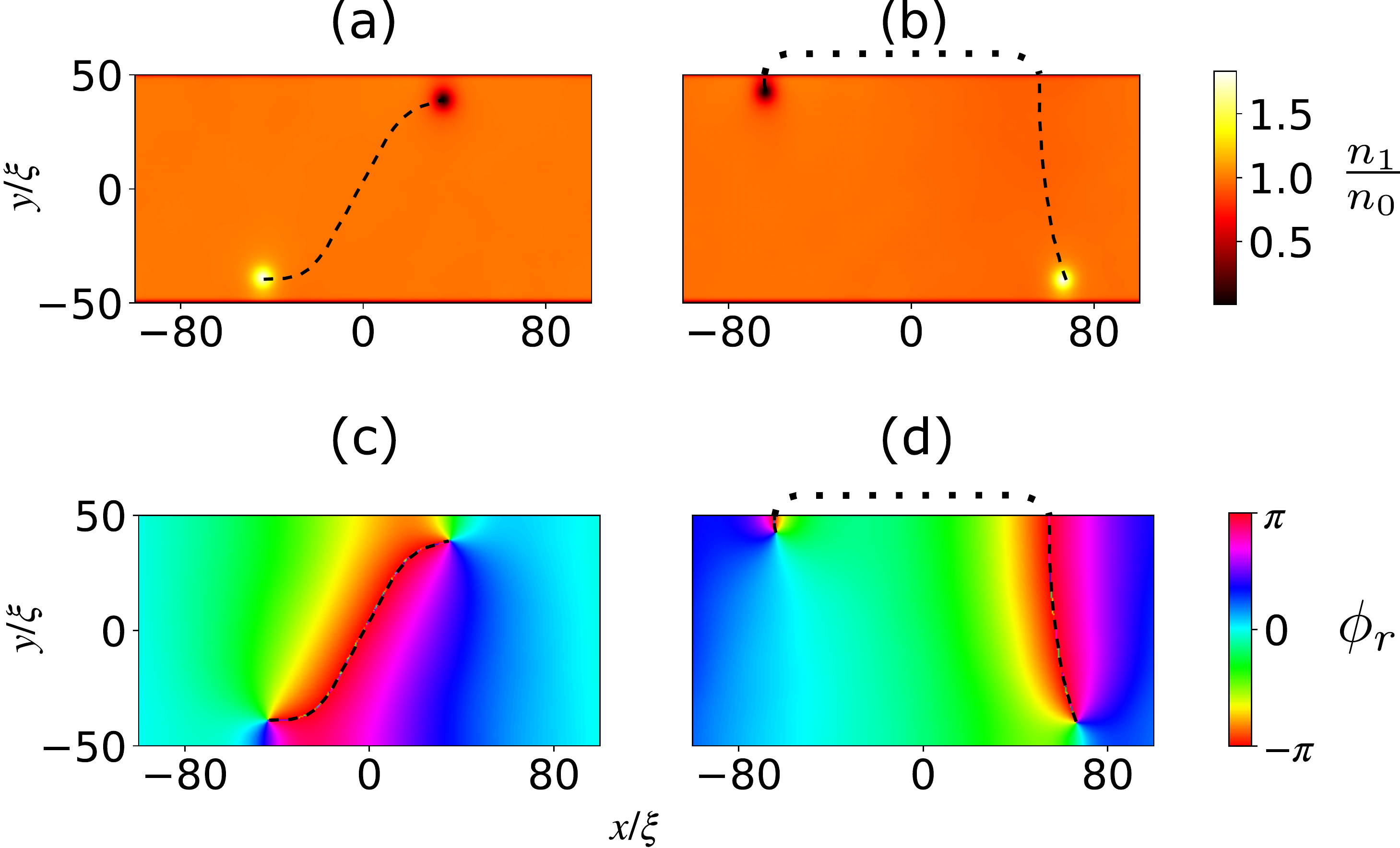}
\caption{Snapshots of the vortex molecule motion in an unbounded trajectory from the time-dependent GPE \eqref{coupledGPE}. The left column (a), (c) shows an early time and the right column (b) and (d) show a later time in the cyan colored trajectory shown in Fig.~\ref{fig:phase_space_g12} (b). The top row shows the color-coded normalized density of component 1, $n_1(\mathbf{r})/n_0$ in panels (a) and (b). The positions of the vortices in component 1 and 2 can be inferred from the bright and dark spots, respectively.  The bottom row shows the relative phase $\phi_r(\mathbf{r})=\arg(\psi_1\psi_2^*)$ in panels (c) and (d). The dashed line in all panels marks the line of $\pi$ phase indicating the presence of a domain wall of the relative phase. The dotted line outside the channel boundaries (on the top and bottom edge) in panels (b) and (d) indicate the topological connection of the domain wall outside of the domain occupied by the superfluid. Parameters are the same as in Fig.~\ref{fig:phase_space_g12}.
}    \label{time_t1}
\end{figure}

In our GPE simulations we have not observed vortex pair production upon stretching vortex molecules. However, we have not seen the boundless growth of domain walls with arbitrary length either. Instead we have seen domain walls bending towards the hard-wall boundaries, where the interaction energy can be contained by routing the vortex filament outside the superfluid. An example of the vortex filament exiting the condensate through the boundary is shown in Fig.~\ref{time_t1} in snapshots taken from the cyan trajectory of Fig.~\ref{fig:phase_space_g12} (b).

So far we have discussed the dynamics of symmetric configurations where the center of mass was at rest in the middle of the channel. Off-setting the center of mass in the $y$ direction leads to an overall translational motion of the vortex molecule in $x$ direction on top of the internal dynamics described above, as expected from the discussion in Sec.~\ref{sec:comseparation}. Additional effects that may be anticipated from the coupling of the relative and center-of-mass degrees of freedom are a distortion of the relative-degree-of-freedom phase space depending of center-of-mass state of motion and vice versa. A deeper study of these effects is beyond the scope of the present work. 
%
% The center-of-mass Hamiltonian $\tilde{\mathcal{H}}_\mathrm{com}$ of Eq.~\eqref{eq:Hcom} 
%

\section{Conclusions} \label{sec:conclusions}

We have set up a point-vortex framework  in which vortex molecule dynamics can 
be explored. 
Applied to the motion of a vortex molecule in a channel geometry we find that the point-vortex model predicts all important qualitative features
of vortex dynamics in the GPE simulation.
The point-vortex model is particularly well suited for inspecting the phase space structure in detail.
It may be interesting to study vortex-molecule dynamics in other geometries, such as billiards, in the future.

Our model could be further refined by taking into account potential inertial effects in the vortex dynamics \cite{Richaud2020,Richaud2021}. 
Such inertial effects may be expected in the case where $g_{12}>0$ due to the partial core filling of a vortex in one component by a density bump in the other. While we have not seen any clear evidence in our simulations, such effects could become more relevant in some situations, e.g.~for imbalanced interaction strengths.

The vortex molecule dynamics in the channel geometry is particularly interesting because it produces unbounded trajectories where the vortex molecule is stretched by a competition of the domain wall tension and vortex attraction from the boundaries. Future work could examine the role quantum fluctuations may play in seeding vortex-antivortex pair creation and thus creating a laboratory analog of color confinement in quantum chromodynamics \cite{etoCollisionDynamicsReactions2020}.

\section*{Acknowledgements} \label{sec:acknowledgements}

We thank Ashton Bradley for discussions and for providing code for vortex detection with \texttt{VortexDistributions.jl} \cite{Bradley2022}.

\appendix

\section{Interaction energy of a vortex molecule}
\label{sec:appendixA}

In order to obtain the total energy of a vortex molecule shown in Fig.~\ref{Interaction_parameter}  we imprint each condensate with a single vortex phase mask at an equal distance $d_\mathrm{ini}/2$ and opposite direction from the center of a square computational domain with dimensions $180\xi \times 180\xi$. We use $d_\mathrm{ini}=60\xi$ in this work. We also locate pinning potentials (peaked Gaussian potentials) on the positions of the phase singularity of each vortex
and evolve the system according to Eq.~\eqref{coupledGPE} in imaginary time until convergence. This creates a vortex molecule with the accurate appropriate phase structure. Then we remove the pinning potential for another round of 
imaginary time evolution during which the molecular distance $d$ changes towards the equilibrium, and plot the energy vs.\ distance. This gives us a fairly accurate picture of the interaction energy as a function of molecular distance $d$. The procedure approximately, but not exactly, produces the minimum energy configuration constrained by the position of the vortex singularities. Indeed, we see small changes in energy values depending on the initial position of the vortex imprint, in particular during early stages of the imaginary time evolution. For this reason we  only use data for fitting the parameterization with $d<40\xi$ when the distance of the initial imprint is $d=60\xi$, as this data is well converged.

\subsection{Twisted real projective plane boundary conditions}

\begin{figure}
\centering
\includegraphics[width=0.5\linewidth,height=0.5\linewidth]{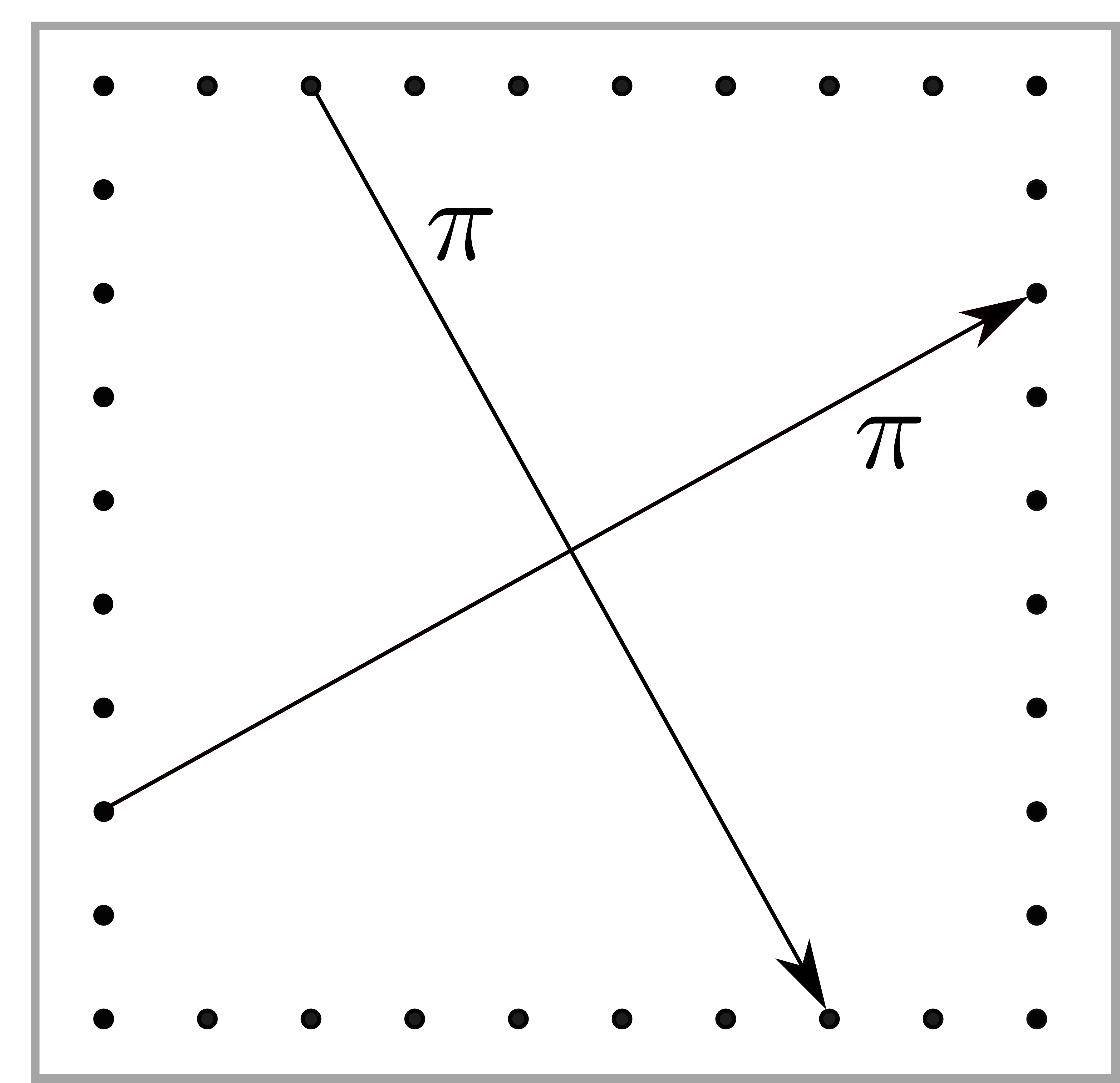}
\caption{Twisted real projective plane boundary conditions.
Grid points on the boundary of the computational domain are connected to diagonally opposite points (by inversion) and restricted to have the same modulus and a complex phase offset of $\pi$. This applies to both complex fields $\psi_1(x,y)$ and $\psi_2(x,y)$.
} \label{gridpoint}
\end{figure}

In order to optimally capture the energy content of a vortex molecule in the absence of boundaries, we use boundary conditions that are designed to approximately generate the density and phase structure expected from a single vortex molecule on an infinite two-dimensional plane. At a distance $r \gg d$ from the vortex molecule, we expect the phase and density structure in each component to be approximately described by that of a simple vortex of Eq.~\eqref{eq:simplevortex}. This will be exact for a vortex molecule with $d=0$. Choosing a square domain and placing the vortex molecule in the center, this implies in particular that the phase of each component has exactly a $\pi$ offset when comparing opposite points on the boundary (by inversion), while the density is the same. Hence we implement boundary conditions that enforce antiperiodicity (i.e.~the same modulus but phase offset by $\pi$) diagonally across the domain. These boundary conditions implementing a real projective plane with a $\pi$ phase twist are illustrated in Fig.~\ref{gridpoint}. Note that the required phase shift of $\pi$ across the diagonal leads to an increased energy cost if the vortex molecule is not centered in the computational domain. Thus imaginary time evolution will automatically center the vortex molecule. 

\begin{figure}
\centering
\includegraphics[width=1\linewidth]{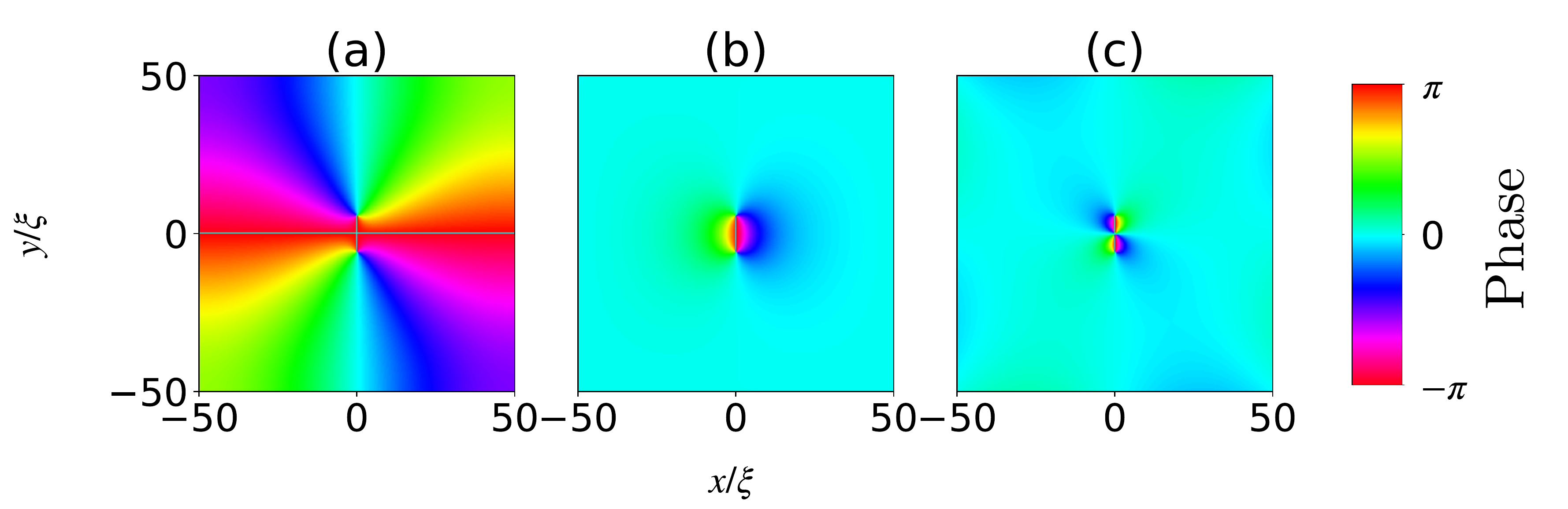}
\caption{Phase structure of a vortex molecule with charge $\kappa=1$ and $d=8.5\xi$ 
under the twisted real projective plane boundary condition.
(a) Total phase of the condensates $\arg(\psi_1 \psi_2)$. 
(b) Relative phase $\arg(\psi_1\psi_2^*)$. 
(c) Residual phase $\arg(\psi_1 \psi_2 / \psi_\mathrm{s}^2)$, where $\psi_\mathrm{s}$ is the complex order parameter of a simple vortex of Eq.~\eqref{eq:simplevortex}, located in the center of the computational domain.
The comparison shows that the total phase of the vortex molecule deviates from that of a simple vortex mainly in a narrowly localised region, with some faint residuals extending about the computational domain. 
Parameters are $g= \mu\xi^2$, $g_{12}=0$, and $\nu=2\times10^{-3}\mu$. 
} \label{VortexMol}
\end{figure}

The phase structure resulting from applying the twisted real projective plane boundary conditions to a charge 1 vortex molecule is shown in Fig.~\ref{VortexMol}. The total phase shown in panel (a) is broadly that of a charge 2 vortex, with the individual unit charges separated at a distance $d$. The relative phase shown in panel (b) reveals the domain wall located between the vortex positions, and healing towards equal phase well before the boundaries are reached. The residual total phase compared to a centered charge 2 vortex shown in panel (c) reveals that most of this residual is localized close to the vortex molecule with length scale $d$. However, faint residuals spanning the whole computational domain can also be distinguished.

\begin{figure}
\centering
\includegraphics[width=1\linewidth,height=0.4\textwidth]{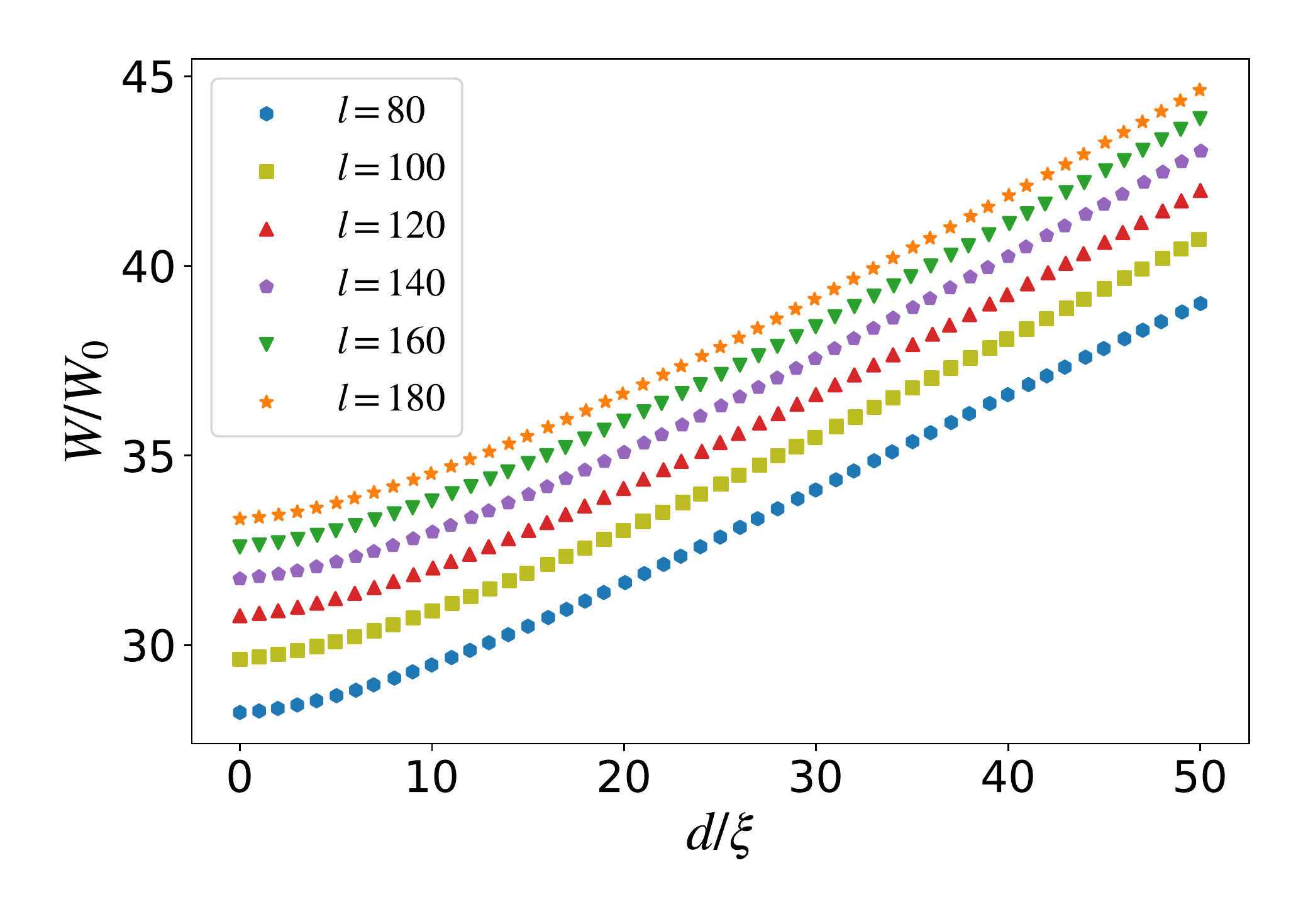}
\caption{Vortex molecule energy as a function of molecular distance $d$ for different sizes $l\xi \times l\xi$ of the computational domain.
Different symbols indicate the different sizes $l$ as indicated in the plot legend. 
The initial distance of vortex seeding is $d_\mathrm{ini}=60\xi$. The slopes vary very little,
but each curve is shifted by a constant due to the additional  energy of the vortex velocity field captured with the
changing size of the computational domain. 
Other parameters are $g= \mu\xi^2$, $g_{12}=0$, and $\nu=2\times10^{-3}\mu$.
} \label{diffbox}
\end{figure}

While Fig.~\ref{VortexMol} mostly supports our assumption that the twisted real projective plane boundaries efficiently remove boundary effects from the simulation, we also repeat the calculation of the vortex molecule energy in computational domains of different size. The results are shown in Fig.~\ref{diffbox}. We see that different box sizes broadly lead to the same energy as a function of molecular distance $d$, but shifted by a constant value. This is expected as a larger computational domain will integrate a larger part of the energy density of the vortex flow pattern, which ultimately is expected to logarithmically diverge with increasing the box size. However, this does not matter for the purpose of Hamiltonian dynamics in the extended point-vortex model of Sec.~\ref{sec:model} where a constant energy offset is irrelevant and does not change the resulting equations of motion. For the smaller box size of $80\xi$ we can see some deviations from the otherwise parallel behavior of the data shown in Fig.~\ref{diffbox}, which we attribute to a boundary effect. It becomes prominent when the molecular separation $d$ is larger than half of the linear box dimension. Hence, we use the data with the largest box size $180\xi \times 180\xi$ for parametrizing the interaction energy.

\subsection{Parameterization}

\begin{figure}
\centering
\includegraphics[width=1\linewidth,height=0.35\textwidth]{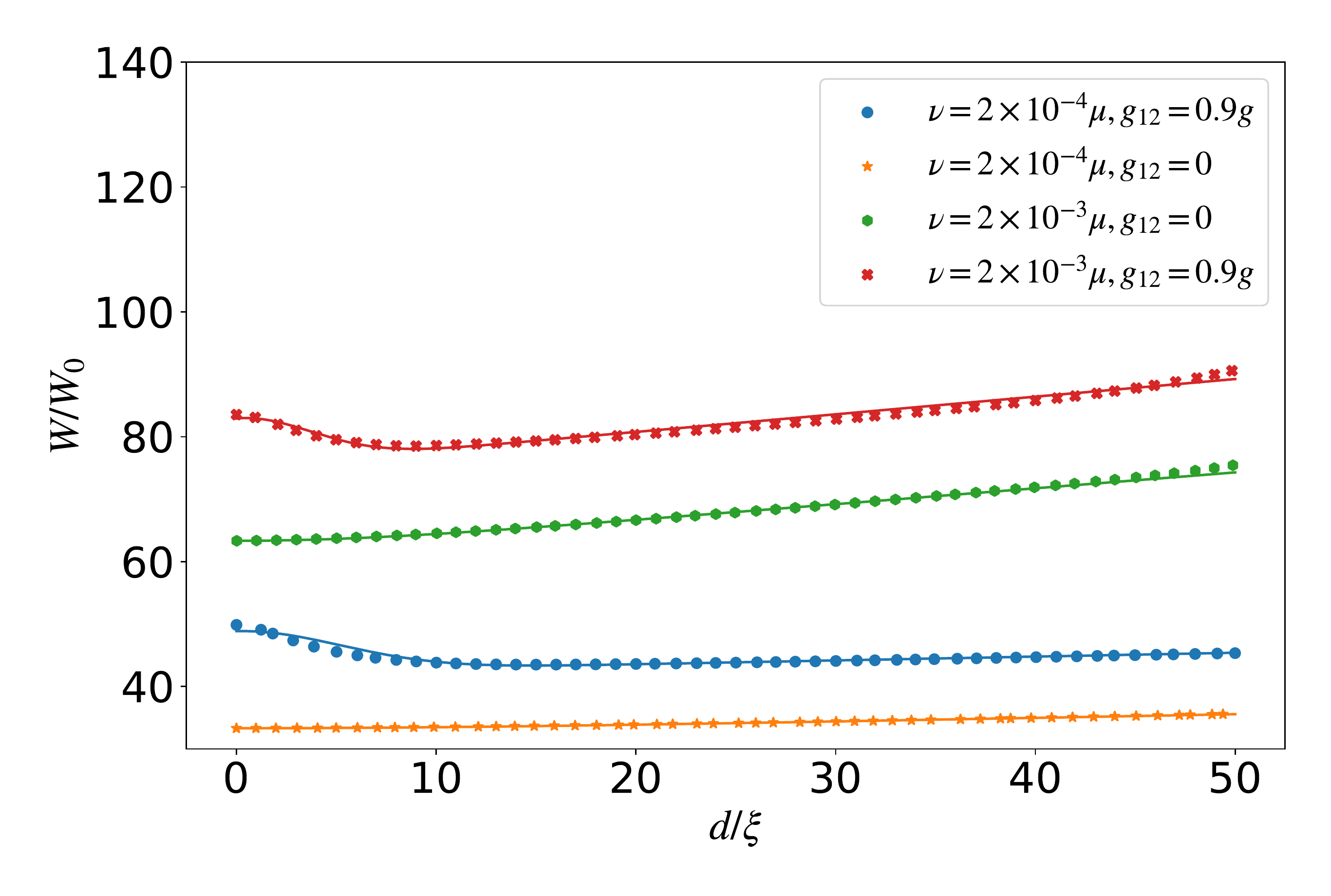}
\caption{Vortex molecule interaction energy and parameterizations for different parameter values. Symbols are numerical data from imaginary time evolution with different values of the 
constant $\nu$ and $g_{12}$ as indicated in the legend. Full lines of the corresponding color are the fits according to Eqs.~\eqref{g12_0_fit} and \eqref{g12_nzero_fit}. Other parameters are $g=0.53 \mu\xi^2$ when $g_{12}=0.9 g$, and $g= \mu\xi^2$ when $g_{12}=0$. The molecular distance is measured in units of healing length $\xi={\hbar}/{\sqrt{m(\mu+\nu)}}$ and 
the energy is measured in $W_0={\hbar^2(\mu+\nu)}/{m(g+g_{12})}$. Energies corresponding to same values of $\nu$ and $g_{12}$ have been shifted by arbitrary amounts for graphical purposes.  
As is evident non-zero $g_{12}$ creates an energy maxima at zero distance due to the absence of core filling. On the other
hand the same values of $\nu$ lead to the same slope for large $d$. Larger $\nu$ results in higher tension from the Josephson vortex and steeper energy slopes. The data shown for $\nu=2\times 10^{-3}\mu$ is the same as shown in Fig.~\ref{Interaction_parameter}.
} \label{Coupling_energy}
\end{figure}

For the purpose of the point vortex model it is very convenient to parameterize the interaction energy of a vortex molecule rather than relying on numerical data that is only available at specific discrete values of the molecular distance $d$.
We have performed calculations of the vortex molecule energy as a function of $d$ for altogether four different parameter values as shown in  Fig.~\ref{Coupling_energy}.

We fit the curves in Fig. \ref{Coupling_energy} with two 
different functional forms depending on the value of $g_{12}$.
For $g_{12}=0$ we use 
\begin{align}
\label{g12_0_fit}
V(d)=& a l_a \log[\cosh(\frac{d}{l_a})]+b l_b\log[\cosh(\frac{d}
    {l_b})] +c  ,
\end{align} 
and for $g_{12}=0.9g$ we use
\begin{align}
\label{g12_nzero_fit}
     V(d)=\alpha e^{- d^2/\beta}+\gamma d + \delta,
\end{align}
where, $a,b,l_a,l_b,\beta,\gamma,\delta$ are fitting parameters. The equilibrium distance of the vortex molecule
is defined as $d_\mathrm{eq}$. This is the molecular distance of the lowest energy configuration. 
We set $\alpha=\beta\gamma \exp(d_\mathrm{eq}^2/\beta)/2 d_\mathrm{eq}$, which ensures that $V(d)$ has a minimum at $d_\mathrm{eq}$. The relevant parameters for both cases are given in Table~\ref{tab:fitparam1} \& \ref{tab:fitparam2}, where $\xi$ is the healing length and $W_0={\hbar^2(\mu+\nu)}/{m(g+g_{12})}$. 

    \begin{table}
    \caption{\label{tab:fitparam1}Fitting parameters for $g_{12}=0.9g$}
        \begin{ruledtabular}
            \begin{tabular}{lllll}
                $\nu/\mu$ & $d_{\mathrm{eq}}/\xi$ & $\beta/\xi^2$ & $\gamma \xi/W_0$ & $\delta/W_0$\\
                \hline
                2 $\times10^{-4} \mu$ &  14.83 & 53.8687 & 0.0614 & 22.3328 \\
                2 $\times10^{-3} \mu$ & 8.83 & 26.9334 & 0.2824 & 25.1213\\
            \end{tabular}
        \end{ruledtabular}
        \caption{\label{tab:fitparam2}Fitting parameters for $g_{12}=0$}
        \begin{ruledtabular}
            \begin{tabular}{llllll}
                $\nu/\mu$ & $a\xi/W_0$ & $l_a/\xi$ & $b\xi/W_0$ & $l_b/\xi$ & c$/W_0$\\
                \hline
                2 $\times10^{-4} $ & 1.0227 & 17.81416077 & -0.96324 & 17.81416075& 33.3192\\
                2 $\times10^{-3} $ & 1.1218  & 10.0701960 & -0.8672 & 10.0701959 & 33.3336\\
            \end{tabular}
        \end{ruledtabular}
    \end{table}
    
This parametrization gives us a form for the interaction energy between the vortex molecules which we use to predict vortex trajectories
along with our analytical model.
     
\bibliography{Solitons,Books,notes} 
%\addcontentsline{toc}{chapter}{Bibliography}

\end{document}